\documentclass[]{emulateapj}
\usepackage{graphicx}
\usepackage[hyperindex,breaklinks]{hyperref}

\def\cf{{\it cf.}}
\def\eg{{\it e.g.}}
\def\etal{{\it et al.}}
\def\etc{{\it etc.}}
\def\ie{{\it i.e.}}

\def\pmb#1{\setbox0=\hbox{$#1$}%
  \kern-0.25em\copy0\kern-\wd0
  \kern.05em\copy0\kern-\wd0
  \kern-0.025em\raise.0433em\box0}

\slugcomment{Revised version, submitted to ApJ, January 13, 2016}

\shorttitle{Bar instability}
\shortauthors{Sellwood}

\begin{document}

\title{Bar instability in disk-halo systems}

\author{J. A. Sellwood}

\affil{Department of Physics and Astronomy, Rutgers University, 136
  Frelinghuysen Road, Piscataway, NJ 08854, USA}

\begin{abstract}
We show that the exponential growth rate of a bar in a stellar disk is
substantially greater when the disk is embedded in a live halo than in
a rigid one having the same mass distribution.  We also find that the
vigor of the instability in disk-halo systems varies with the shape of
the halo velocity ellipsoid.  Disks in rigid halos that are massive
enough to be stable by the usual criteria, quickly form bars in
isotropic halos and much greater halo mass is needed to avoid a strong
bar; thus stability criteria derived for disks in rigid halos do not
apply when the halo is responsive.  The study presented here is of an
idealized family of models with near uniform central rotation and that
lack an extended halo; we present more realistic models with extended
halos in a companion paper.  The puzzle presented by the absence of
strong bars in some galaxies having gently rising inner rotation
curves is compounded by the results presented here.
\end{abstract}



\section{Introduction} \label{sec:intro}

It has long been known that the dominant dynamical instability of
massive stellar disks leads to the formation of a strong bar
\citep{MPQ70, Hohl71, Kaln72, Kaln77, Jala07}.  The mode is an
exponentiating standing wave between two reflection points: corotation
where traveling waves are strongly amplified as they swing from
leading to trailing \citep{GLB65, JT66}, and the disk center where an
ingoing trailing wave reflects into an outgoing leading wave.
\citet{Toom81}, who elucidated this mechanism for the bar mode, also
argued that the instability could be quelled in at least two ways.
Toomre's preferred method was to interrupt feedback through the
center, perhaps by ensuring that the waves are damped at an inner
Lindblad resonance \citep[ILR,][]{Mark74}, which should be present if
the galaxy has a dense bulge, for example.  This feature accounts for
the remarkable linear stability of the Mestel, and other self-similar,
disks \citep{Zang76, ER98}.  High-quality simulations \citep{Sell89,
  SM99, SE01} of massive disks having dense centers do indeed avoid
bar-formation, as predicted by linear theory, but an ultra-responsive
disk can amplify even a modest level of noise to such an extent that
the linear theory expectation of damping at resonances can be
overwhelmed, and particles become trapped into a bar through a
non-linear instability.  Thus, this stabilizing method may not always
be effective, especially in a galaxy that experiences a strong
perturbation.  Random motions make the disk less responsive, and
generally contribute to stability \citep[\eg][]{AS86}.

It is indeed true that the bar instability is quelled when the disk is
immersed in a massive, unresponsive spheroidal component.  The effect
of the extra central attraction is to shorten the preferred wavelength
of gravitationally-driven disturbances in the disk until bi-symmetric
disturbances become no more than mild kinematic waves that can no
longer extract much amplification from the global potential well of
the galaxy \citep{Toom81, BT08}.  This is the basis of several
well-known stability criteria \citep{OP73, ELN82, CST95}.

However, these criteria assume the halo is unresponsive, and here we
show that a responsive halo encourages the bar instability in disks,
and that these stability criteria are inadequate when the disk is
immersed in a live halo.  A hint of a more vigorous bar instability
may have been present in the simulations reported by \citet{AM02} and
by \citet{MVSH06}, but the focus of those papers was on the surprising
secular growth of the bars, which grew to encompass most of the disk.
Here we find that, for models of isolated disk galaxies that do not
have a high central density, the disk-halo system possesses a unstable
bar-forming mode that remains quite vigorous unless the disk has very
low mass or the velocity distribution of the halo is strongly radially
biased.

In order to develop an understanding of the instability, we focus on
the linear modes of highly idealized models of unstable disks that
start out axisymmetric and form bars, which is clearly a situation
that is unlikely to arise in nature.  In addition, the halos of our
models are not at all extended, a choice that enabled the use of
anisotropic velocity distributions and also excluded the complications
caused by secular bar growth.  \citet{SB15} present slightly more
realistic simulations having extended halos in which disk mass rises
through slow infall and find that models lacking a central mass
concentration continue to manifest similar behavior.

\section{Model set up} \label{sec.setup}

We wish to create equilibrium, axisymmetric, disk-halo models with no
bulge, in which the shape of the velocity ellipsoid of the halo
particles can be varied. 

\subsection{Halo and disk}

The requirement that the halo distribution function (DF) be
anisotropic is particularly limiting, since many available methods to
create equilibrium halos in the presence of a disk assume an isotropic
halo DF.  However, if we choose a spherical halo for which anisotropic
DFs are known, we can use the adiabatic invariance of actions to
determine the equilibrium DF of the halo after an additional mass
component has built up, as originally set out by \citet{Youn80}.
\citet{SM05} gave a full description of the method for halo
compression and showed that a spherical approximation is entirely
adequate, even when the compressing component is a disk that is
heavier than those used here.  If the DF of the uncompressed model was
originally a function of just two integrals, the DF of the compressed
spherical halo remains a function of two actions: the radial action
and total angular momentum.

We require a spherical model with known anisotropic DFs that also
lacks a density cusp.  From this very limited selection \citep{BT08},
we have chosen the family of anisotropic models given by
\citet{Dejo87} for the Plummer sphere, which has the density profile
\begin{equation}
\rho(r) = {3M_P \over 4\pi a^3}\left[1 + \left({r \over
    a}\right)^2\right]^{-5/2}.
\label{eq.Plum}
\end{equation}
This mass distribution has a harmonic core, consistent with the gently
rising rotation curves in late-type galaxies of lower luminosity
\citep[\eg][]{CF85, Cati06, THINGS}.  Since the density declines as
$r^{-5}$ when $r \gg a$, where $a$ is the core radius, the function
could not represent an extended halo.  However, the models allow us to
study the linear stability of a disk embedded in a live spherical
component whose velocity anisotropy can be varied.  It clearly cannot
capture secular angular momentum exchanges between bars and extensive
halos \citep{Wein85, DS00}; this behavior is included in experiments
described in a later paper \citep{SB15}.

\citet{Dejo87} gives expressions (in terms of hypergeometric
functions) for the DF, characterized by a parameter $q$ that is
related to the usual anisotropy parameter \citep{BT08} as:
\begin{equation}
\beta \equiv 1 - {\sigma_\phi^2 \over \sigma_r^2} = 1 - {\sigma_\theta^2
  \over \sigma_r^2} = {qr^2 \over 2(a^2+r^2)}.
\end{equation}
The DF is positive or zero for all bound values of the specific
energy, $E$, and allowed specific total angular momentum, $L$, when
$-\infty < q \leq 2$.  These halos, which have no net angular
momentum, are isotropic when $q=0$, radially biased when $q>0$ and
tangentially biased when $q<0$, with $q=-\infty$ giving a sphere
composed of entirely circular orbits.  All models with finite $q$
become isotropic as $r/a \rightarrow 0$; anisotropy increases rapidly
outwards and is already half its asymptotic value by $r=a$.  The
maximal radial bias is for $q=2$, where $\beta \rightarrow 1$ for $r
\gg a $.  We use just three different halo DFs in this work: $q=0$
(isotropic), $q=2$ (maximal radial bias), and $q=-15$ (strong
tangential bias).

Simulations of an isolated Plummer sphere with $q=2$ revealed that it
suffers from a radial orbit instability.  The growth rate, $\sim
0.015(GM_P/a^3)^{1/2}$, was low but the outer halo became strongly
prolate by $t\sim 300(a^3/GM_P)^{1/2}$ in a model with $10^6$
particles.  We continue to use initially spherical models that are
compressed by the disk, and allow them to evolve self-consistently.

When compressed by a disk, the spherical density profile is changed,
of course, and \citet{SM05} used this family of anisotropic models to
illustrate how compressibility of a halo varied with radial bias,
finding, as expected, that halo compression was reduced as radial
pressure rises.  The models in this paper have lower disk masses,
however, and compression has a minor effect on the density and a
negligible effect on the shape of the velocity ellipsoid.

We insert an exponential disk of surface density
\begin{equation}
\Sigma(R) = {M_d \over 2\pi R_d^2} e^{-R/R_d}.
\end{equation}
We limit the extent of the disk by tapering the surface density with a
function that varies as a cubic polynomial from unity at $R=4.5R_d$ to
zero at $R=5R_d$.  The disk has a Gaussian density profile in the
vertical direction with $z_0=0.1R_d$.

This family of disk-halo models therefore has two further parameters,
the ratio of halo to disk mass, $M_P/M_d$, and ratio of the disk scale
length $R_d$ to the halo core radius $a$.  We have experimented with
halo masses in the range $1 \leq M_P/M_d \leq 5$ and keep $a = 1.5R_d$
for the reason given below.

\subsection{Initial velocities}

To make the halo finite, we adopt a maximum energy of halo particles
equal to the potential in the disk plane at $r = 8a$ and select halo
particles with energies less than this from the compressed halo DF in
the manner described by \citet[][Appendix A]{DS00}.  The rapid density
decline when $r \gg a$ (eq.~\ref{eq.Plum}) implies that just a few
percent of the total mass is eliminated by our energy bound, and the
halo is in excellent equilibrium in the combined potential of the disk
and halo.

We set the disk particles in orbital motion, add random speeds such
that Toomre's $Q=1.5$, and use the Jeans equations in the local
approximation to set initial in-plane and vertical velocity balance
for the disk particles in the numerically-computed gravitational field
of the total mass distribution.  As the disk mass generally
contributes a moderate fraction at most of the central attraction, the
Jeans equations yield a disk that is also in excellent initial
equilibrium.

\begin{figure}[t]
\begin{center}
\includegraphics[width=.9\hsize]{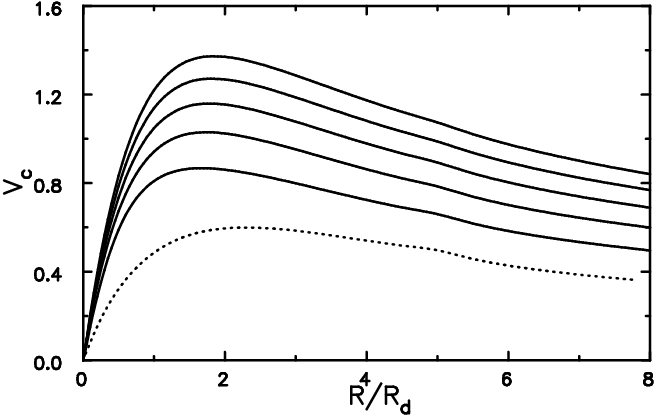}
\end{center}
\caption{The radial variation of the circular speed in a series of
  models with differing halo masses.  The unchanging disk contribution
  is shown by the dotted line, while the full drawn curves are for
  $M_P/M_d = 1$, 2, 3, 4, and 5.}
\label{fig.rotcur}
\end{figure}

Figure~\ref{fig.rotcur} shows the rotation curves of five of our
models with differing halo masses.  The dotted curve shows the disk
contribution computed from the finitely thick and tapered disk, which
is the same for all cases.  The shapes of the rotation curves are
little affected by the mass of the halo, which was the reason for
choosing $a = 1.5R_d$.

\citet{Sack97} defined a maximum disk as one for which
$\left(V_d/V_{\rm tot}\right)^2 \ga 0.85$ at the radius at which the
disk contribution peaks.  The models shown in Fig.~\ref{fig.rotcur}
are all therefore strongly submaximal since, at $R=2R_d$, they have
$\left(V_d/V_{\rm tot}\right)^2 = 0.476$, 0.336, 0.264, 0.219, and
0.187 respectively for $M_P/M_d = 1$, 2, 3, 4, and 5.

The decrease of the circular speed at large radii is a consequence of
our adopted Plummer halo.  While some results in this study would
probably be changed were a more extensive halo to be used, as we note
in places, it seems ikely that a supporting response to the linear
instability that forms the initial bars (the principal result
presented below) would be provided by any reasonable live halo.  In
fact, the absence of an extended halo is an advantage, since it
enables us to study the dynamical instability that forms the bar,
without the complication of secular bar growth and braking caused by
dynamical friction with an extended halo.

All the rotation curves rise nearly linearly from zero at the center,
and this lack of shear in the inner parts is highly unfavorable for
spiral arm development by swing amplification, as \citet{SC84}
reported.  Thus our strongly submaximal models in rigid halos do not
manifest multi-arm spirals of the kind found in other simulations of
submaximal disks with more general rotation curve shapes
\citep[\eg][]{Fuji11, GKC12}.

Henceforth, we adopt $M_d$ as our unit of mass, $R_d$, as our unit of
length and set $G=1$.  Consequently, a dynamical time is
$(R_d^3/GM_d)^{1/2}$ and the velocity unit is $(GM_d/R_d)^{1/2}$.
Since we have smaller disk galaxies, such as M33, in mind, we suggest
a scaling to physical units that sets $R_d = 1.4\;$kpc \citep{RV94}
and $t_0 = 15\;$Myr, which implies $M_d = 2.7 \times
10^{9}\;$M$_\odot$ and velocities scale as $(GM/R_d)^{1/2} \simeq
91\;$km/s.  Thus the velocity rise to the peak shown in
Fig.~\ref{fig.rotcur} roughly matches the inner rotation curve of M33
\citep{Kam15}.

\subsection{Numerical procedure}

We use the hybrid $N$-body code described by \citet[][Appendix
  B]{Sell03}, in which the gravitational field is computed with the
aid of two grids: a 3D cylindrical polar grid for the disk component,
and a spherical grid for the halo component with a surface harmonic
expansion on each grid shell.  A comprehensive description of our
numerical methods is given in an on-line manual \citep{Sell14}.

We generally use $10^6$ particles to represent each component, a time
step $\delta t = 0.01$ dynamical times, include sectoral harmonics
$0\leq m \leq 8$ in the force determination on the cylindrical polar
grid, where the cubic spline softening \citep{Mona92} length is
$0.2R_d$, and surface harmonics $0 \leq l \leq 4$ on the spherical
grid.  Aside from a time offset caused by a different seed amplitude,
we found the evolution was unaffected when the numbers of disk and
halo particles were varied over the range $10^5 \leq N \leq 10^7$, but
$N=10^4$ was clearly inadequate.  We also verified that moderate
changes to the time step and grid size have little effect on the
outcome of the simulations.

We recentered the grids on the particle centroid every 16 time steps,
since an offset between the particle distribution and the center of
our polar grids could lead to numerical artifacts.  As a further
check, we reran one case using a Cartesian grid, which has no
preferred center and has adequate resolution for these models that do
not have a high central density, and once again the evolution followed
the same path.

\section{Bar growth}
\subsection{Halo anisotropy}

\begin{figure}[t]
\begin{center}
\includegraphics[width=.9\hsize]{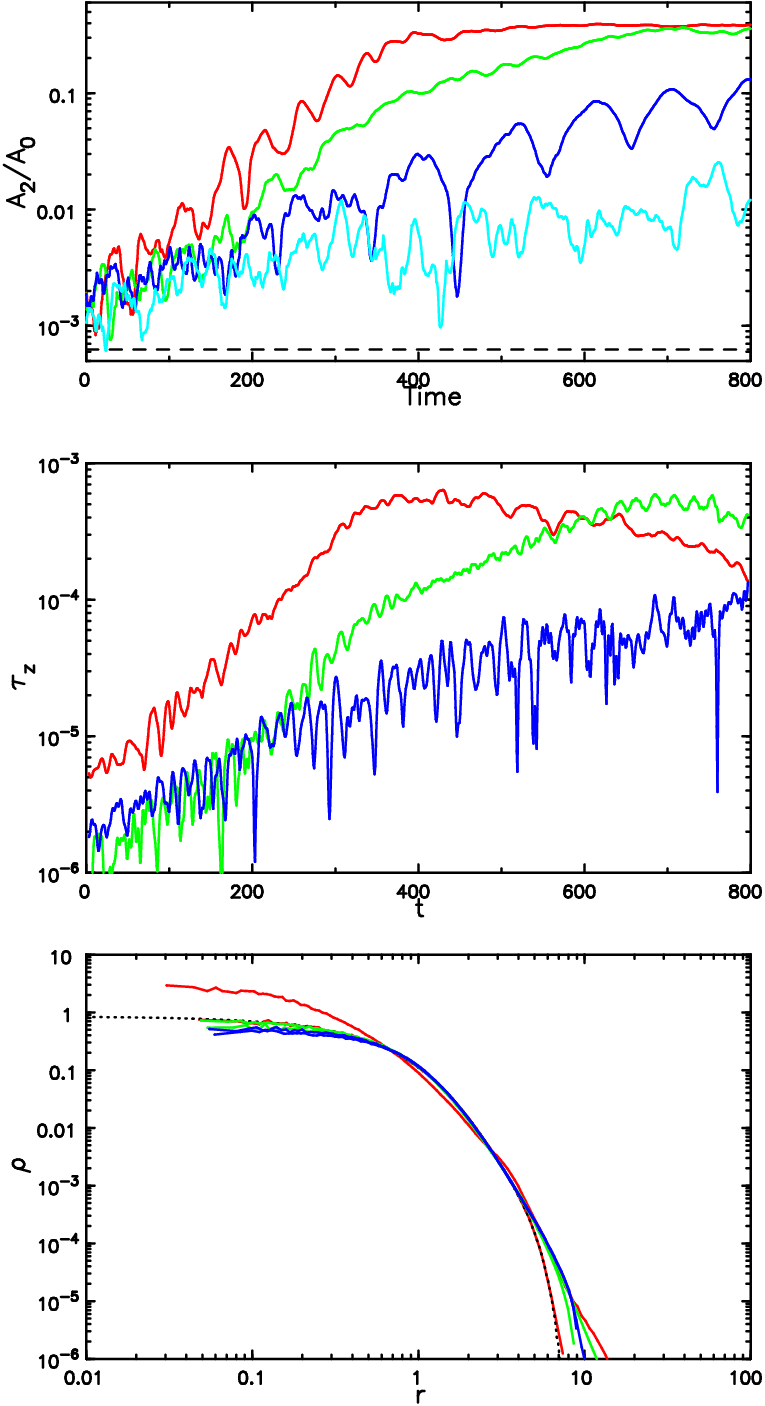}
\end{center}
\caption{The evolution of four models with almost identical mass
  distributions, with $M_P/M_d=3$, but having different DFs for the
  halo.  The top panel shows the evolution of the bar amplitude in the
  disk component; the red curve is for an tangentially biased halo DF,
  the green for an isotropic halo, the blue for a radially biased halo
  DF, and the cyan for a rigid halo.  The dashed horizontal line is
  the expected value of the ordinate if the particles were distributed
  at random in a axisymmetric disk.  The middle panel shows the time
  evolution of the torque exerted by the disk on the live halo,
  colored as in the top panel.  The bottom panel shows the initial and
  final halo density profiles for the three live halo models.}
\label{fig.baramp}
\end{figure}

The top panel in Figure~\ref{fig.baramp} shows the evolution of the
bar amplitude in the disk in four models having almost identical mass
distributions, but having different DFs for the halo.  The quantity
plotted is the ratio $A_2/A_0$, where
\begin{equation}
A_m(t) = \left| \sum_j \mu_je^{im\phi_j} \right|,
\end{equation}
where $\mu_j$ is the mass and $\phi_j(t)$ the cylindrical polar angle
of the $j$-th particle at time $t$, and the summation includes only
disk particles.  A straight line in this log-linear plot would be the
signature of exponential growth; periodic modulations may indicate
beats between two waves rotating with different pattern speeds.

The disk embedded in a rigid halo, cyan line, possesses very slowly
growing, coherent bi-symmetric instabilities (see \S\ref{sec.modes}),
causing the value of $A_2/A_0$ to rise slowly, by a factor $\sim 10$
over the duration of the simulation.  However, the bar amplitude grows
more rapidly in all three cases with live halos, and the behavior
differs remarkably with the nature of the halo DF.  The disk bar
becomes strongest in the tangentially biased halo ($q=-15$, red curve),
while slower bar growth occurs for the radially biased ($q=2$, blue
curve) halo DF.  The isotropic halo ($q=0$, green curve) is
intermediate.

Notice that the value of $A_2/A_0$ reaches $\sim 0.4$ for the
tangentially biased (red) and isotropic (green) models.  These two
bars become extremely strong inside $R\la3R_d$, where density
variations from peak to trough are almost 100\%.  The value of
$A_2/A_0 \ga 0.1$ for the bar when the halo is radially biased halo
(blue) is still visually discernible as an oval distortion to the
inner disk by $t=800$.  This mass model should be comfortably stable
in a rigid halo by conventional stability criteria: $t_{\rm OP} =
0.10$ \citep{OP73} and $V = 1.16(GM/R_d)^{1/2}$ at the peak of the
rotation curve \citep[Fig.~\ref{fig.rotcur}][]{ELN82}.  This
prediction is indeed borne out by the cyan line in the top panel of
Fig.~\ref{fig.baramp} (see also \S\ref{sec.stab}), but the same model
is clearly unstable when the halo is live whatever the shape of the
halo velocity ellipsoid.

The middle panel shows the (smoothed) time evolution of the torque
exerted by the disk on the halo, determined from the rate of gain of
$L_z$ by the halo, for the three cases with the live halo.  Comparison
of these curves with those in the top panel, reveals that the bar
grows by giving angular momentum to the halo, since both the bar
amplitude and the halo torque grow exponentially at approximately the
same rate.

Note that the radially biased halo, which suffered from a radial orbit
instability when isolated, seemed to remain closely spherical when the
disk was inserted.  Furthermore, since the bar grew more slowly, there
is no evidence from the evolution of this model that a possible
instability in the halo encourages the formation of a disk bar.

The loss of angular momentum from the disk causes the disk mass
distribution to become more concentrated, which in turn causes
additional compression of the halo, overwhelming any tendency for the
halo to expand as a result of gaining angular momentum.  Halo
compression (bottom panel) is most pronounced for the tangentially
biased DF, red curves, partly because more angular momentum is lost
from the disk (middle panel) and partly because the lower radial
pressure makes an tangentially biased halo more compressible.  The
smaller angular momentum changes and stiffer halos make the halo
density changes in the other two models too small to notice.

\subsection{Other halo masses}

\begin{figure}[t]
\begin{center}
\includegraphics[width=.9\hsize]{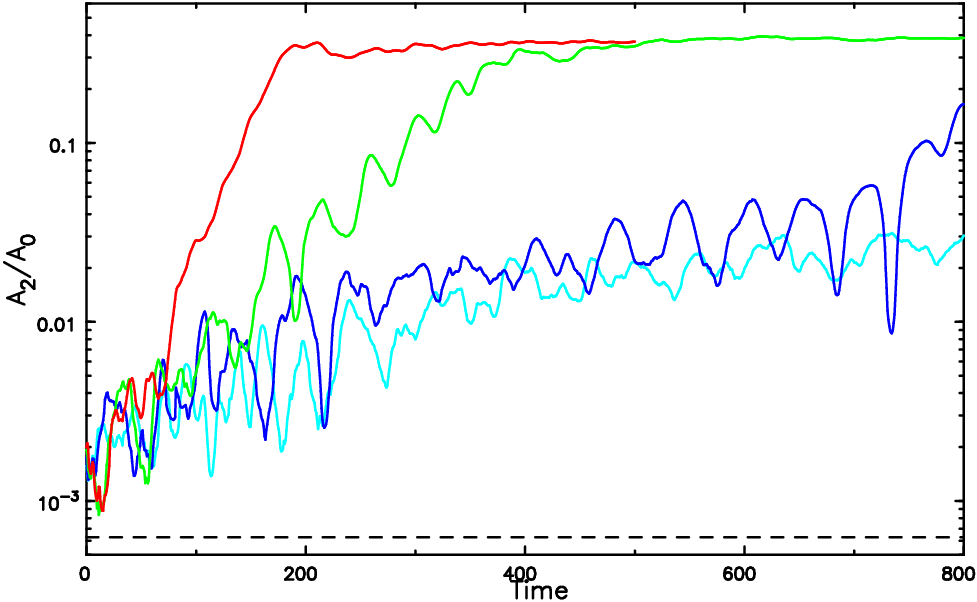}
\includegraphics[width=.9\hsize]{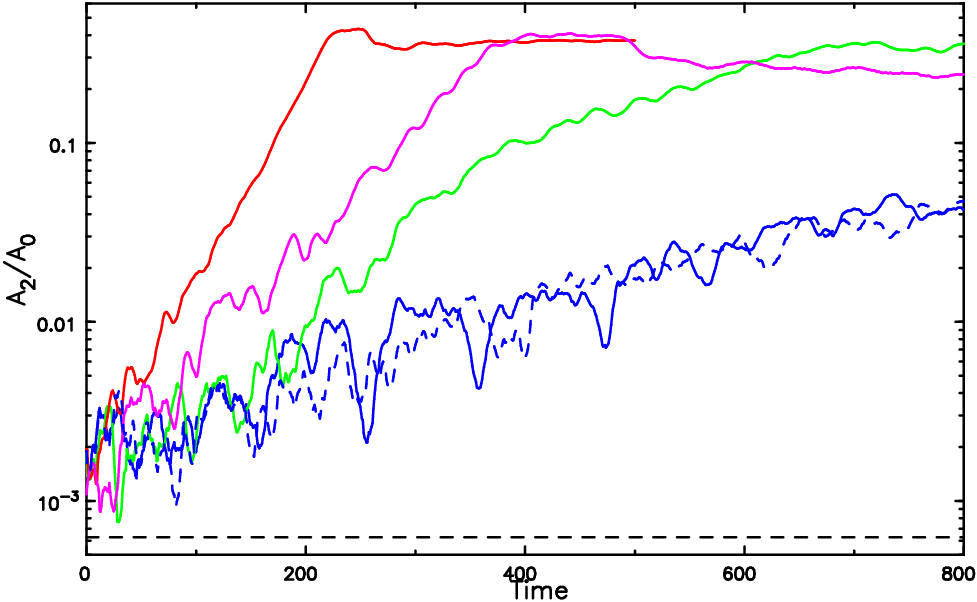}
\end{center}
\caption{The time evolution of the bar amplitude in models with
  differing $M_P/M_d$.  In the upper panel, all halos have tangentially
  biased DFs, while the halo DFs are isotropic in the lower panel.
  The value of $M_P/M_d$ is denoted by the line color: red for
  $M_P/M_d = 2$, green for $M_P/M_d = 3$ (reproduced from
  Fig~\ref{fig.baramp}), blue for $M_P/M_d = 4$, and cyan for $M_P/M_d
  = 5$.  The magenta line in the lower panel is for a rigid halo with
  $M_P/M_d = 2$, and the dashed blue line is from the same $M_P/M_d =
  4$ model, but with the evolution computed on a Cartesian grid.}
\label{fig.hmass}
\end{figure}

Figure~\ref{fig.hmass} shows the consequence of changing the halo
mass.  The results in the upper panel are from simulations in which
the halo velocity ellipsoid was strongly tangentially biased ($q=-15$),
while it was isotropic in the lower panel.  Since the disk in a rigid
halo, even, has a vigorous bar instability when $M_P/M_d = 2$ (magenta
line, lower panel), the rapid rise in the bar amplitude for the
corresponding live halos, red lines, is hardly surprising, although it
is noteworthy that the growth rates are clearly higher when the halo
is live, and more so for the tangentially biased case.

However, the other curves for more massive halos continue to reveal
bar growth on a time scale that slows both with increasing halo mass
and with a shrinking fraction of near circular orbits in the halo.
The green curves are from models with $M_P/M_d = 3$ that are simply
reproduced from Fig.~\ref{fig.baramp}, but the blue curves and the
cyan curve are from models with $M_P/M_d = 4$ and 5, respectively.
While these models have lower growth rates, the instability is still
clearly present.

\subsection{Properties of the bars} \label{sec.bars}

Some bars in these disks become very strong, and buckle out of the
plane, in the usual manner, to make peanut shapes that are very
pronounced for the strongest bars.  The bars always form and remain
fast, \ie\ corotation is close to the end of the bar.  The halo mass
beyond the disk edge at $R=5R_d$ in these simulations is less than 8\%
of the total in all the runs reported in this paper.  The absence of
an extended halo almost certainly inhibits bar slow-down through
dynamical friction, and therefore these bars should not be regarded as
counter-examples to the expectation \citep{DS00} that ${\cal R} \equiv
a_B/R_c \ga 1.4$ for strong bars immersed in dense halos.\footnote{
  The dimensionless ratio ${\cal R} \equiv R_c / a_B$, originally
  defined by \citet{Elme96}, with $R_c$ being the bar corotation
  radius and $a_B$ being the semi-major axis of the bar, characterizes
  the bar angular speed.}  This aspect also prevented the kind of
secular growth reported by \citet{AM02} and \citet{MVSH06}, where the
bars grew continuously to encompass the entire original disk.

\section{Linear instabilities} \label{sec.modes}
\subsection{Mode fitting}

Although these are not ``quiet start'' simulations, the initial noise
is low enough that the disturbances grow almost 100-fold before
saturating, which affords reasonable estimates of the exponential
growth rate.  We used the mode fitting apparatus described in
\citet{SA86} to estimate eigenfrequencies for the most rapidly growing
modes over the period of linear growth.  The Fourier coefficients of
the density on the cylindrical polar grid, as well as logarithmic
spiral transforms of the disk particles yield data that may be fitted
with exponentially growing and rotating modes.  The fitted mode
generally has a complex frequency $\omega$, where $\Re(\omega) =
m\Omega_p$, with $\Omega_p$ being the pattern speed and $m=2$ for
bisymmetric modes, and $\Im(\omega)$ is the $e$-folding rate.

\begin{figure}[t]
\begin{center}
\includegraphics[width=.9\hsize]{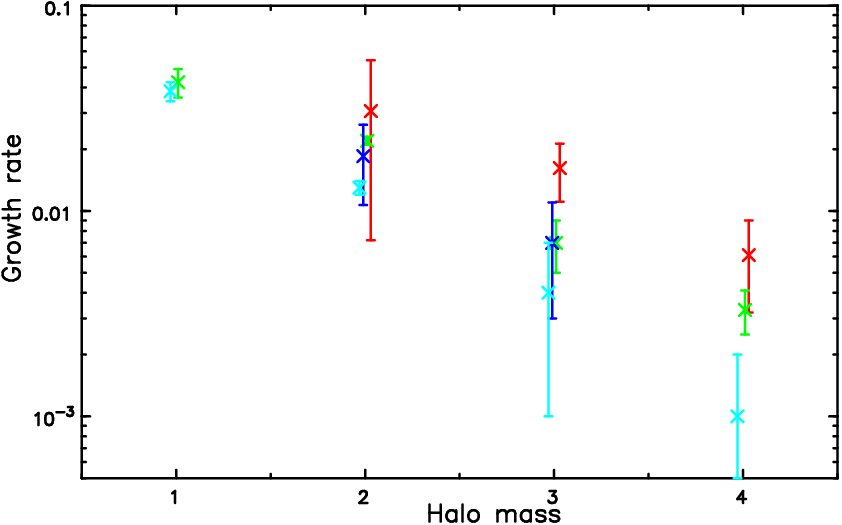}
\end{center}
\caption{The estimated growth rates of the most unstable mode as a
  function of increasing halo mass.  The colors signify the nature of
  the halo DF, as in Fig.~\ref{fig.baramp}; red for tangentially
  biased, green for isotropic, blue for radial bias, and cyan for a
  frozen halo.  The halo mass is the same for each group of points,
  which are offset horizontally for clarity.}
\label{fig.modes}
\end{figure}

The crosses in Figure~\ref{fig.modes} show the best estimate of the
growth rate of the most rapidly growing mode in each simulation, and
the error bars indicate the spread in values from multiple fits to
slightly different time ranges, fitted data type, rescaling factors,
\etc \ More than a single mode seemed to be required to fit the data
from some simulations, especially where the amplitude appears to be
modulated.  In these cases, the estimated pattern speeds of the two
disturbances differ by about the right amount to account for the beat
period, but a single, strong bar prevails in every case that reaches
large amplitude.

Note that all the measured pattern speeds of all the growing modes
reported here are high enough to avoid ILRs.

The case with the rigid halo with $M_{\rm P}/M_d = 3$ (cyan curve, top
panel of Fig.~\ref{fig.baramp}), possesses perhaps three very mildly
growing coherent bisymmetric disturbances that we were able to
identify in this long simulation.  The shapes of these three waves
show the classic pattern \citep{Toom81, BT08} of interference between
ingoing and outgoing waves, with more nodes for the lower frequency
waves, as expected for a low mass disk in a rigid halo.

The behavior of the estimated growth rates in Fig.~\ref{fig.modes}
with both halo mass and with the halo velocity distribution is
consistent with the evolution of the amplitudes presented in
Figs~\ref{fig.baramp} and \ref{fig.hmass}.  At all halo masses, the
cyan points indicate that disks in rigid halos have the lowest growth
rates compared with those from live halos, whatever the form of the
DF.  Notice that the spread in growth rate at each halo mass broadens
as the halo becomes more massive, implying that growth rates decline
more slowly with increasing halo mass in live halos and most slowly in
halos with tangentially biased DFs.  Compared with that in the rigid
halo, the bar growth rates range up to five times higher for the most
massive responsive halo; even a isotropic DF causes the mode to have a
three times higher growth rate.

\subsection{Mode mechanism}

Phase coherence and exponential growth indicate that the disturbances
are dynamical instabilities of the disk-plus-halo system.  Since the
modes lack ILRs, the instability appears to have a similar feed-back
mechanism to that of the bar instability of massive disks reviewed in
the introduction, with the exception that amplification is boosted by
interaction with the halo.

It is well-known \citep{LBK72} that, when halo interactions are
excluded, a disk inside corotation loses angular momentum, and that
outside gains, as a self-excited mode grows.  The outward transfer of
angular momentum across corotation is accomplished by the spiral
torque of the disturbance density.  In our simulations with live
halos, the torque from the disk on the halo (Fig.~\ref{fig.baramp}
middle panel) grows exponentially with the mode until it saturates,
indicating that the bar is excited in these cases by the loss of
angular momentum to the halo, and probably to the outer disk also.
The substantial differences in the disk bar growth rate when the halo
DF is changed, while keeping the halo density almost constant, is
clear evidence that halo particles on near circular orbits couple most
strongly to the density disturbance in the disk.

Increasing the halo mass slows the instability.  The relative
frequencies of disk and halo orbits must be little affected by
changing halo mass, since both the disk and halo particles move in the
same global potential.  The more likely explanation is that the
relatively less massive and more tightly wrapped disturbances in the
disk couple less effectively to the more massive halos.

\section{Discussion} \label{sec.discuss}
\subsection{Two obvious questions}

The results presented here are so startling, it is reasonable to ask
two questions: (1) can they be right? and, if so, (2) why is the
importance of responsive halos to the bar instability not more
generally known?

We already reported in \S\ref{sec.setup} that our results were
insensitive to particle number, time step, grid resolution and even
the type of grid, as shown by the solid and dashed blue lines in the
lower panel of Fig.~\ref{fig.hmass}.  Furthermore, \citet{AM02} and
\citet{MVSH06}, who both used direct-$N$ methods, stressed the effect
of the halo on the disk bar in their simulations of sub-maximal disks,
but the bars in their models must have resulted from the same linear
instability that we have identified.  Thus there seems little room to
doubt the numerical results reported here.

The well-cited papers by \citet{AM02} and \citet{MVSH06} reported just
a couple of cases that may well have been unstable had they used rigid
halos, yet the formation of bars in sub-maximal disks has not
hitherto been followed up with a systematic study.  In part, this
could be because to use a tree code for this kind of study would
require $\ga 100$ times more computer time \citep{Sell14} than the
code we used here, which is able to compute the evolution of these
models for a Hubble time with modest computational resources.  Yet the
more likely reason for no previous focused investigation may be that
it is unfashionable to use simulations to study a single aspect of
galaxy dynamics.  Instead, many workers prefer to include all possible
relevant complications, such as ``gastrophysics,'' halo triaxiality,
substructure, \etc\ into every simulation.  While a strong case can
be made for this approach, it comes with two clear disadvantages: it
not only makes the simulations very costly in computer time, but it
also renders the few results that can be obtained much more difficult
to understand.

\subsection{Disk stability} \label{sec.stab}

The difference of behavior between live and rigid halos is not
qualitative, but simply that growth rates are higher.  As has been
reported before \citep[\eg][]{Toom81}, disks in even very massive
rigid halos continue to possess mild bisymmetric instabilities, even
though they are believed to be ``stable.''  

Thus, the practical question of disk stability is perhaps best cast as
whether a bar would grow by less than some large factor, one hundred
or $\sim 5\;e$-folds say, in a Hubble time or 800 dynamical times,
which translates to $\Im(\omega) \la 0.006$ in the units of this
paper.  By this criterion, disk stability requires a more massive live
halo than a rigid one.  Judging from the slopes of the growth rate
measurements in Fig.~\ref{fig.modes}, the halo need be some 30\% --
50\% more massive for the growth rate in a responsive isotropic halo
to match that in a rigid halo.

Note that this conclusion is indicated by isotropic halos in this one
mass model, and that a more general stability criterion, if one
exists, would require an extensive study to explore many different
mass models and halo DFs.  Note also, that in models having a dense
center, such as a bulge, the disk can be linearly stable at much
higher mass fraction than in these cases with gently rising rotation
curves \citep{SB15}.

\subsection{Implications for real galaxies}

Higher growth rates for bar instabilities in live halos make the
survival of unbarred disks in bulgeless galaxies, which already had no
clear explanation \citep{Sell13}, still more perplexing.  The growth
rate can be suppressed by reducing the disk mass, or by making the halo
DF strongly radially biased, or by raising the level of random motion
in the disk.  We discuss all these options in turn, finding each quite
unattractive.

The usual argument against the rigid halo explanation for disk
stability is that it predicts \citep{SC84, ABP87} that galaxy disks
should be dominated by small-scale, multi-arm spirals, as found by
others \citep[\eg][]{Fuji11, GKC12, DVH13}.  We have shown here that
live halos, unless they are strongly radially biased right to near the
center, would require still lower disk masses.  In further
experiments, to be described later, we again observe tightly-wrapped,
multi-arm spirals in cool disks in compressed, isotropic Einasto halos
that are massive enough to suppress the bar.  However, nature does not
appear to prefer this solution, since observations
\citep[\eg][]{Davi12}, find that the majority of spiral patterns in
galaxies have $m=2$ or 3.

Relaxed halos from dark matter only $\Lambda$CDM simulations are
mostly isotropic near their centers and become mildly radially biased
by the virial radius, where $\beta \simeq 0.35$ \citep{DM11}.  Very
little seems to have been reported of how the velocity ellipsoid shape
is affected by baryonic infall and feedback, but a dramatic change
would seem unlikely.  Thus there is no reason to expect a {\em strong}
radial bias in the inner parts of the halos of real galaxies that
would materially improve the stability of real galaxy disks.

Disks with large degrees of random motion will be more stable than
cool disks \citep{AS86}.  Current galaxy formation models
\citep{Gove10, Scan12, Chri14} favor strong ``feedback'' from dense
gas clumps that impulsively eject large gas masses from the disk after
adiabatic infall.  The effect of this repeated irreversible cycle is
both to lower the halo density and to make the remaining stellar disk
quite puffy with large random motions in all three components.  While
random motion among the disk stars must be stabilizing, the galaxies
that present the greatest stability puzzles are late-type spirals with
stellar masses in the range few $\times 10^8$ to several $\times
10^9\;$M$_\odot$.  These galaxies have little in the way of bulge
light, which is why their rotation curves rise slowly in the inner
parts \citep{Cati06}, and lie in the mass range of greatest flattening
\citep[\eg][]{SJM10}.  Furthermore, they usually have reasonably
well-developed spiral patterns, which is another indicator that the
disk cannot be so hot as to be dynamically unresponsive.

Thus, if galaxies are indeed embedded in responsive dark matter halos,
then the fact that a significant fraction of them lack a strong bar is
still more perplexing than before.

It is particularly puzzling that M33 supports a coherent two-arm
spiral pattern and at the same time has no more than a very weak bar
\citep{RV94}.  This is because a bi-symmetric spiral pattern suggests
a massive disk; strong swing amplification is generally required
either for a self-excited mode or for an amplified mild tidal
perturbation.  Since swing amplification at $m=2$ is negligible in a
low-mass disk \citep{Toom81}, a much stronger tidal field would be
required to produce a pronounced two-arm distortion.  However, the
velocity map in the inner parts \citep{Kam15} seems too regular to be
consistent with the strong tidal perturbation a low-mass disk would
require.

\section{Conclusions} 

We have demonstrated, for model galaxies lacking a dense center, that
a disk embedded in a live, \ie\ responsive, halo tends to form a strong
bar.  This is true, even when the disk is sufficiently sub-maximal
that it is expected to be stable by the traditional stability criteria
\citep{OP73, ELN82}.  It is important to realize that this instability
is one of the combined disk and halo system; therefore, stability
studies of disks embedded in rigid halos do not tell the full story,
and stability criteria derived from them do not apply to galaxies
having live halos.

We have found that none of the unstable modes presented in this work
has an ILR, and that the mechanism for the linear mode remains that of
feedback through the center, as outlined by \citet{Toom81}.  However,
his picture of swing amplification at corotation cannot be the whole
story, since the work reported here demonstrates that angular momentum
loss to the halo is another source of wave amplification.  In other
work \citep{SB15}, we find that linear bar instabilities are absent in
models with dense bulges that inhibit wave reflection off the center.

The vigor of the disk instability in a live halo is strongly related
to the ability of disk density variations to couple to near circular
orbits in the halo.  Therefore, compared with an isotropic halo of a
given mass, higher growth rates are found when the halo DF is
tangentially biased, and lower when it is radially biased.  Note that
in all cases presented here, the halo has no net rotation.  Increasing
the halo mass slows the instability, but slowly growing modes persist
even in very strongly submaximal disks embedded in isotropic halos.

The bars formed in these simulations are very strong, but their
lengths are consistent with those observed in real galaxies.
\citet{Gado11} finds the median value $a_B/R_d \simeq 1.5 \pm 0.5$,
where $a_B$ is the bar semi-major axis.  Note that the value of $R_d$ he
used was that in the galaxy he observed, \ie, after the bar formed.
We find $a_B/R_{d,\rm new} \simeq 1.5$ in our models, since the scale
length of the disk outside the bar is about twice that of the original
disk.  The abruptly truncated halos of the models used here prevented
further growth of the bars, of the kind reported by \citet{AM02} and
by \citet{MVSH06}, which would make them unrealistically long and
strong \citep[\cf][]{Erwi05}.  Such bars would be unmistakable in any
well resolved galaxy image, and their absence in galaxies with gently
rising rotation curves presents a major unsolved puzzle.

Galaxy dynamics has turned out to be much richer and more complex than
anyone imagined.  The fact that a surprise such as that presented here
could turn up after almost a half-century of simulations should remind
us of how much more we have to learn, and how much may lie buried in
the confusing behavior of simulations that attempt to include
everything that could be important.

\acknowledgments 

I wish to thank Joel Berrier for his persistent reports of bars in
simulations with live halos that prompted this study.  His help, both
with discussions and for researching the topic of anisotropy in halos
from structure formation simulations, is gratefully acknowledged.  The
comments of an anonymous referee helped to clarify several points.
This work was supported by NSF grants AST/1108977 and AST/1211793.


\begin{thebibliography}{}

\def\aap{A\&A}
\def\aj{AJ}
\def\apj{ApJ}
\def\apjl{ApJL}
\def\apjs{ApJS}
\def\apss{Ap.\ Sp.\ Sci.}
\def\araa{ARAA}
\def\jcop{J. Comp.\ Phys.}
\def\mnras{MNRAS}
\def\newa{New. Astron.}
\def\PhD{PhD.\ thesis}
\def\nat{Nature}
\def\pf{{\it Phys.\ Fluids}}
\def\PhD{{\it PhD thesis}}
\def\phya{{\it Physica\/} A}
\def\rpp{Rep.\ Prog.\ Phys.}
\def\Omit#1{{ \etal}}

\bibitem[\protect\citeauthoryear{Athanassoula \etal}{1987}]{ABP87}
Athanassoula, E., Bosma, A. \& Papaioannou, S. 1987, \aap, {\bf 179}, 23

\bibitem[\protect\citeauthoryear{Athanassoula \& Misiriotis}{2002}]{AM02}
Athanassoula, E. \& Misiriotis, A. 2002, \mnras, {\bf 330}, 35

\bibitem[\protect\citeauthoryear{Athanassoula \& Sellwood}{1986}]{AS86}
Athanassoula, E. \& Sellwood, J. A. 1986, \mnras, {\bf 221}, 213

\bibitem[\protect\citeauthoryear{Berrier \& Sellwood}{2016}]{SB15}
Berrier, J. C. \& Sellwood, J. A. 2016, \apj, (in preparation)

\bibitem[\protect\citeauthoryear{Binney \& Tremaine}{2008}]{BT08}
Binney, J. \& Tremaine, S. 2008, {\it Galactic Dynamics\/} (2nd ed.; Princeton: Princeton University Press)

\bibitem[\protect\citeauthoryear{Carignan \& Freeman}{1985}]{CF85}
Carignan, C. \& Freeman, K. C. 1985, \apj, {\bf 294}, 494

\bibitem[\protect\citeauthoryear{Catinella, Giovanelli \& Haynes}{2006}]{Cati06}
Catinella, B., Giovanelli, R. \& Haynes, M P. 2006, \apj, {\bf 640}, 751

\bibitem[\protect\citeauthoryear{Christensen \etal}{2014}]{Chri14}
Christensen, C. R., Governato, F., Quinn, T.,\Omit{ Brooks, A. M., Shen, S., McCleary, J., Fisher, D. B. \& Wadsley, J.} 2014, \mnras, {\bf 440}, 2843

\bibitem[\protect\citeauthoryear{Christodoulou, Shlosman \& Tohline}{1995}]{CST95}
Christodoulou, D. M., Shlosman, I. \& Tohline, J. E. 1995, \apj, {\bf 443}, 551

\bibitem[\protect\citeauthoryear{Davis \etal}{2012}]{Davi12}
Davis, B. L., Berrier, J. C., Shields, D. W.,\Omit{ Kennefick, J., Kennefick, D., Seigar, M. S., Lacy, C. H. S. \& Puerari, I.} 2012, \apjs, {\bf 199}, 33

\bibitem[\protect\citeauthoryear{Debattista \& Sellwood}{2000}]{DS00}
Debattista, V. P. \& Sellwood, J. A. 2000, \apj, {\bf 543}, 704

\bibitem[\protect\citeauthoryear{de Blok \etal}{2008}]{THINGS}
de Blok, W. J. G., Walter, F., Brinks, E., Trachternach, C., Oh, S.-H. \& Kennicutt, R. C., Jr.\ 2008, \aj, {\bf 136}, 2648

\bibitem[\protect\citeauthoryear{Dejonghe}{1987}]{Dejo87}
Dejonghe, H. 1987, \mnras, {\bf 224}, 13

\bibitem[\protect\citeauthoryear{Diemand \& Moore}{2011}]{DM11}
Diemand, J. \& Moore, B. 2011, ASL {\bf 4}, 297

\bibitem[\protect\citeauthoryear{D'Onghia, Vogelsberger \& Hernquist}{2013}]{DVH13}
D'Onghia, E., Vogelsberger, M. \& Hernquist, L. 2013, \apj, {\bf 766}, 34

\bibitem[\protect\citeauthoryear{Efstathiou, Lake \& Negroponte}{1982}]{ELN82}
Efstathiou, G., Lake, G. \& Negroponte, J. 1982, \mnras, {\bf 199}, 1069

\bibitem[\protect\citeauthoryear{Elmegreen}{1996}]{Elme96}
Elmegreen, B. 1996, in IAU Colloq.\ {\bf 157}, {\it Barred Galaxies}, ed.\ R. Buta, D. A. Crocker \& B. G. Elmegreen (San Francisco: ASP Conf series {\bf 91}), 197

\bibitem[\protect\citeauthoryear{Erwin}{2005}]{Erwi05}
Erwin, P. 2005, \mnras, {\bf 364}, 283

\bibitem[\protect\citeauthoryear{Evans \& Read}{1998}]{ER98}
Evans, N. W. \& Read, J. C. A. 1998, \mnras, {\bf 300}, 106

\bibitem[\protect\citeauthoryear{Fujii \etal}{2011}]{Fuji11}
Fujii, M. S., Baba, J., Saitoh, T. R., Makino, J., Kokubo, E. \& Wada, K. 2011, \apj, {\bf 730}, 109

\bibitem[\protect\citeauthoryear{Gadotti}{2011}]{Gado11}
Gadotti, A. D. 2011, \mnras, {\bf 415}, 3308

\bibitem[\protect\citeauthoryear{Goldreich \& Lynden-Bell}{1965}]{GLB65}
Goldreich, P. \& Lynden-Bell, D. 1965, \mnras, {\bf 130}, 125

\bibitem[\protect\citeauthoryear{Grand, Kawata \& Cropper}{2012}]{GKC12}
Grand, R. J. J., Kawata, D. \& Cropper, M.  2012, \mnras, {\bf 426}, 167

\bibitem[\protect\citeauthoryear{Governato \etal}{2010}]{Gove10}
Governato, F., Brook, C., Mayer, L.,\Omit{ Brooks, A., Rhee, G., Wadsley, J., Jonsson, P., Willman, B., Stinson, G., Quinn, T. \& Madau, P.} 2010, \nat, {\bf 463}, 203

\bibitem[\protect\citeauthoryear{Hohl}{1971}]{Hohl71}
Hohl, F. 1971, \apj, {\bf 168}, 343

\bibitem[\protect\citeauthoryear{Jalali}{2007}]{Jala07}
Jalali, M. A. 2007, \apj, {\bf 669}, 218

\bibitem[\protect\citeauthoryear{Julian \& Toomre}{1966}]{JT66}
Julian, W. H. \& Toomre, A. 1966, \apj, {\bf 146}, 810

\bibitem[\protect\citeauthoryear{Kalnajs}{1972}]{Kaln72}
Kalnajs, A. J. 1972, \apj, {\bf 175}, 63

\bibitem[\protect\citeauthoryear{Kalnajs}{1977}]{Kaln77}
Kalnajs, A. J. 1977, \apj, {\bf 212}, 637

\bibitem[\protect\citeauthoryear{Kam \etal}{2015}]{Kam15}	
Kam, Z. S., Carignan, C., Chemin, L., Amram, P. \& Epinat, B. 2015, \mnras, {\bf 449}, 4048

\bibitem[\protect\citeauthoryear{Lynden-Bell \& Kalnajs}{1972}]{LBK72}
Lynden-Bell, D. \& Kalnajs, A. J. 1972, \mnras, {\bf 157}, 1

\bibitem[\protect\citeauthoryear{Mark}{1974}]{Mark74}
Mark, J. W-K. 1974, \apj, {\bf 193}, 539

\bibitem[\protect\citeauthoryear{Martinez-Valpuesta, Shlosman \& Heller}{2006}]{MVSH06}
Martinez-Valpuesta, I., Shlosman, I. \& Heller, C. 2006, \apj, {\bf 637}, 214

\bibitem[\protect\citeauthoryear{Miller, Prendergast \& Quirk}{1970}]{MPQ70}
Miller, R. H., Prendergast, K. H. \& Quirk, W. J. 1970, \apj, {\bf 161}, 903

\bibitem[\protect\citeauthoryear{Monaghan}{1992}]{Mona92}
Monaghan, J. J. 1992, \araa, {\bf 30}, 543

\bibitem[\protect\citeauthoryear{Ostriker \& Peebles}{1973}]{OP73}
Ostriker, J. P. \& Peebles, P. J. E. 1973, \apj, {\bf 186}, 467

\bibitem[\protect\citeauthoryear{Regan \& Vogel}{1994}]{RV94}
Regan, M. W. \& Vogel, S. N. 1994, \apj, {\bf 434}, 536

\bibitem[\protect\citeauthoryear{Sackett}{1997}]{Sack97}
Sackett, P. D. 1997, \apj, {\bf 483}, 103
	
\bibitem[\protect\citeauthoryear{S\'anchez-Janssen, M\'endez-Abreu \& Aguerri}{2010}]{SJM10}
S\'anchez-Janssen, R., M\'endez-Abreu, J. \& Aguerri, J. A. L. 2010 \mnras, {406}, L65

\bibitem[\protect\citeauthoryear{Scannapieco \etal}{2012}]{Scan12}
Scannapieco, C., Wadepuhl, M., Parry, O. H.,\Omit{ Navarro, J. F., Jenkins, A., Springel, V., Teyssier, R., Carlson, E., Couchman, H. M. P., Crain, R. A., Dalla Vecchia, C., Frenk, C. S., Kobayashi, C., Monaco, P., Murante, G., Okamoto, T., Quinn, T., Schaye, J., Stinson, G. S., Theuns, T., Wadsley, J., White, S. D. M. \& Woods, R.} 2012, \mnras {\bf 423}, 1726

\bibitem[\protect\citeauthoryear{Sellwood}{1989}]{Sell89}
Sellwood, J. A. 1989, \mnras, {\bf 238}, 115

\bibitem[\protect\citeauthoryear{Sellwood}{2003}]{Sell03}
Sellwood, J. A. 2003, \apj, {\bf 587}, 638

\bibitem[\protect\citeauthoryear{Sellwood}{2013}]{Sell13}
Sellwood, J. A. 2013, in {\it Planets Stars and Stellar Systems}, v.{\bf 5}, eds.\ T. Oswalt \& G. Gilmore (Heidelberg: Springer) p.~923 (arXiv:1006.4855)

\bibitem[\protect\citeauthoryear{Sellwood}{2014}]{Sell14}
Sellwood, J. A. 2014, arXiv:1406.6606 (on-line manual: \hfil\break {\tt http://www.physics.rutgers.edu/$\sim$sellwood/manual.pdf})

\bibitem[\protect\citeauthoryear{Sellwood \& Athanassoula}{1986}]{SA86}
Sellwood, J. A. \& Athanassoula, E. 1986, \mnras, {\bf 221}, 195

\bibitem[\protect\citeauthoryear{Sellwood \& Carlberg}{1984}]{SC84}
Sellwood, J. A. \& Carlberg, R. G. 1984, \apj, {\bf 282}, 61

\bibitem[\protect\citeauthoryear{Sellwood \& Evans}{2001}]{SE01}
Sellwood, J. A. \& Evans, N. W. 2001, \apj, {\bf 546}, 176

\bibitem[\protect\citeauthoryear{Sellwood \& McGaugh}{2005}]{SM05}
Sellwood, J. A. \& McGaugh, S. S. 2005, \apj, {\bf 634}, 70

\bibitem[\protect\citeauthoryear{Sellwood \& Moore}{1999}]{SM99}
Sellwood, J. A. \& Moore, E. M. 1999, \apj, {\bf 510}, 125

\bibitem[\protect\citeauthoryear{Toomre}{1981}]{Toom81}
Toomre, A. 1981, in {\it The Structure and Evolution of Normal Galaxies}, ed.\ S. M. Fall \& D. Lynden-Bell (Cambridge: Cambridge University Press), p.~111

\bibitem[\protect\citeauthoryear{Weinberg}{1985}]{Wein85}
Weinberg, M. D. 1985, \mnras, {\bf 213}, 451

\bibitem[\protect\citeauthoryear{Young}{1980}]{Youn80}
Young, P. 1980, \apj, {\bf 242}, 1232

\bibitem[\protect\citeauthoryear{Zang}{1976}]{Zang76}
Zang, T. A. 1976, \PhD, MIT

\end{thebibliography}
\end{document}